\documentclass[english,runningheads,letter]{llncs}[2018/03/10]

\usepackage[ngerman,main=english]{babel}

\usepackage{epsfig,color,inputenc,amsthm,paralist,algorithmic,url,float}
\usepackage{booktabs} % For formal tables
\usepackage{paralist} 
\usepackage{listings}
\usepackage{graphicx}
\usepackage{fancyvrb}
\usepackage[normalem]{ulem}
\usepackage{xcolor,soul}
\usepackage{balance}
\usepackage{amssymb}
\usepackage{tabularx}
\usepackage{multirow}
\usepackage{booktabs}
\usepackage{threeparttable}
\usepackage{mathtools}
\usepackage{subcaption}
\usepackage[numbers]{natbib}
\usepackage{makecell}
\usepackage{url}

\newcommand{\NAME}{MPD }

\begin{document}

\title{MPD: Moving Target Defense through Communication Protocol Dialects}
%If Title is too long, use \titlerunning
%\titlerunning{Short Title}

%Single insitute
\author{Yongsheng Mei, Kailash Gogineni, Tian Lan \and Guru Venkataramani}
%If there are too many authors, use \authorrunning
%\authorrunning{First Author et al.}
\institute{The George Washington University \\ \email{$\{$ysmei, kailashg26, tlan, guruv$\}$@gwu.edu}}

%% Multiple insitutes - ALTERNATIVE to the above
% \author{%
%     Firstname Lastname\inst{1} \and
%     Firstname Lastname\inst{2}
% }
%
%If there are too many authors, use \authorrunning
%  \authorrunning{First Author et al.}
%
%  \institute{
%      Insitute 1\\
%      \email{...}\and
%      Insitute 2\\
%      \email{...}
%}

\maketitle

\begin{abstract}
Communication protocol security is among the most significant challenges of the Internet of Things (IoT) due to the wide variety of hardware and software technologies involved. Moving target defense (MTD) has been adopted as an innovative strategy to solve this problem by dynamically changing target system properties and configurations to obfuscate the attack surface. Nevertheless, the existing work of MTD primarily focuses on lower-level properties (e.g., IP addresses or port numbers), and only a limited number of variations can be generated based on these properties. In this paper, we propose a new approach of MTD through communication protocol dialects (MPD) - which dynamically customizes a communication protocol into various protocol dialects and leverages them to create a moving target defense. Specifically, \NAME harnesses a dialect generating function to create protocol dialects and then a mapping function to select one specific dialect for each packet during communication. To keep different network entities in synchronization, we also design a self-synchronization mechanism utilizing a pseudo-random number generator with the input of a pre-shared secret key and previously sent packets. We implement a prototype of \NAME and evaluate its feasibility on standard network protocol (i.e., File Transfer Protocol) and internet of things protocol (i.e., Message Queuing Telemetry Transport). The results indicate that \NAME can create a moving target defense with protocol dialects to effectively address various attacks - including the denial of service attack and malicious packet modifications - with negligible overhead.


\end{abstract}

\section{Introduction}
\label{sec:intro}

The Internet of Things (IoT) refers to the concept of a large number of smart objects and devices connected to the Internet, offering diverse capabilities, such as sensing, actuating, processing, and communication. It integrates and relies on various enabling components, e.g., software, application libraries, middleware, embedded systems, and network artifacts. Any vulnerability in these components would lead to exploitations to create serious threats - such as the denial of service attack and reconnaissance attack- to the IoT system.

The Moving Target Defense (MTD) has been developed as an effective defense strategy to dynamically (and randomly) change the properties of configurations of a target system while maintaining its essential functionalities to diversify its defense mechanism and obfuscate the resulting attack surfaces. It significantly increases the work factor of an adversary to launch an effective attack towards a constantly evolving defense. System attributes (and thus the potential attack surfaces) that can be dynamically changed to confuse attackers include instruction sets, address space layouts, IP addresses, port numbers, proxies, virtual machines, and operating systems~\cite{cho2020toward}. Existing work on MTD has primarily focused on low-level attributes, such as instruction set randomization~\cite{barrantes2003randomized,kc2003countering} and address space layout randomization~\cite{hund2013practical,seibert2014information}. Some other MTD methods target network-level features, such as IP address randomization~\cite{al2011toward,jafarian2012openflow}, virtualization-based MTD~\cite{okhravi2012creating} and software-defined networking based MTD~\cite{macfarland2015sdn,wang2017random}. However, considering the security of communication protocols in IoT, these MTD methods cannot achieve desired defense diversity against potential attacks, as the low-level protocol properties to be mutated (e.g., IP addresses or port numbers) are minimal. 

In this paper, we propose a new approach for communication protocol MTD 
through protocol dialects, denoted by MPD. Our key idea of MPD is to automatically create many protocol dialects, which are variations of the target protocol created by mutating its handshake and message formats while keeping the communication functionalities unchanged. By selecting different protocol dialects and switching between them on the fly, we craft an MTD solution with substantially boosted diversity in communication. In order to enable easy management of dialects, our proposed solution also leverages distributed hash and keyspace partitioning to allow adding, removing, modifying any protocol dialect independent of others in MPD. We show that MPD requires very low overhead and is suitable for lightweight communication protocols in IoT using client-server architecture, e.g., File Transfer Protocol (FTP) and Message Queuing Telemetry Transport (MQTT). These protocols are usually lightweight and focus on data exchange efficiency while offering very limited security mechanism~\cite{andy2017attack, nebbione2020security}. We argue that vigilantly customizing protocol dialects for higher-level features such as handshake and message format is crucial for directly protecting IoT security from potential attacks, especially for insecure communication channels in IoT that adversaries frequently target.

At the core of creating customized packet dialects, the critical problem is to design appropriate dialect generating functions that are easy to implement and will not introduce a high computational cost. Manually creating dialect templates is costly and also unrealistic for a huge program. Apart from that, we need to consider how to make the automatically created dialect applied on each packet vary following a random and unpredictable pattern. Only by increasing the uncertainty of the moving target can our approach help improve communication security. We propose two types of dialect generating functions - targeting at handshake and message format, respectively - to automatically create a number of protocol dialects as candidates for MTD. During communication, different dialects will be selected by a mapping function and applied to every handshake to dynamically change the defense properties, thus varying the attack surface on the fly. Furthermore, to deal with potential disruptions and attacks on the communication channel, we design the self-synchronization mechanism motivated by the self-synchronizing stream cipher~\cite{millerioux2010self} to ensure that the server and clients in an IoT system always return to a synchronized state after a limited error propagation.

\vspace{0.1in}
The main contributions of our work are as follows: 

\begin{itemize}
	\item We propose \NAME, an automated and self-synchronous framework for creating and leveraging protocol dialects for effective MTD in IoT communications. Given a specific protocol, \NAME can automatically generate enough dialects per user needs and apply them to certain packets/handshakes during communication to improve security.
	
	\item \NAME leverages distributed hash and keyspace partition for easy dialect management. It also implements a self-synchronization mechanism. Packets are cached and used in conjunction with a pseudo-random generator to ensure the randomness and unpredictability of dialects.
	
	\item Evaluation using two communication protocols, namely FTP and IoT protocol MQTT, shows that \NAME can efficiently generate a large number of dialects and dynamically select the dialects for a round of communication for improved security in IoT systems.
	
\end{itemize}

\section{Motivation}
\label{overview}

In this paper, we focus on IoT communication protocols that adopt a server-client architecture and employ packets for a communication, e.g., FTP and MQTT. Such protocols often emphasize communication efficiency and offer limited protection and security. We consider a threat model in which adversaries in an IoT system can dominate all communications, erase, replay, and replace arbitrary control packets between the client and server. In particular, an attacker can launch the Denial of Service (DoS) attack to exhaust the resources available at a victim network entity (e.g., a server or a client). This type of attack can be easily implemented by replaying the system with numerous connection requests~\cite{haripriya2019secure} or intentionally keeping all the connections alive on the network busy~\cite{vaccari2020slowite}. Besides, in our threat model, we also consider malicious modification of communication. Given a pre-owned privilege, an attacker can modify genuine control packets or inject malformed control packets to launch the command injection attack~\cite{liu2017vehicle,vuong2015decision}, and this could undermine the system's integrity by maliciously disrupting handshakes or guiding a victim to execute unauthorized code. These attacks can also be indicated by the documented real-world vulnerabilities such as CVE-2019-9760 on FTP and CVE-2016-10523 on MQTT. We note that the above-mentioned threat models support a proof-of-concept for \NAME as a promising new direction of MTD in communication security without being over-aiming for an all-encompassing solution. Therefore, this work serves as an initial step toward a comprehensive solution based on MTD through protocol dialects.

Moving target defense (MTD) \cite{nitrd2013iwg} enables us to create, analyze, evaluate, and deploy mechanisms and strategies that are diverse and that continually shift and change over time to increase complexity and cost for attackers, limit the exposure of vulnerabilities and opportunities for attacks, and increase system resiliency. Current MTD methods are limited in increasing complexity against potential attacks because low-level system attributes (e.g., IP addresses and port numbers) only offer a limited degree of freedom for mutation. Due to the above reasons, we design \NAME, an MTD technique combined with protocol dialect that leverages more application-layer properties of communication protocols and efficiently increases the number of variations that we can create.

The goal of \NAME is to customize a standard protocol and create a moving target defense by (i) fabricating desired protocol dialects and (ii) allowing servers/clients to dynamically select dialects for communication in a synchronous fashion. We first introduce our definition of protocol dialects as follows:

\begin{definition}{\textbf{Protocol Dialect.}}\label{def:dialect}
	Given a standard communication protocol, a dialect in this paper is a variation created by mutating its packets and handshakes while keeping the communication functionalities unchanged. More precisely, let $\mathcal{D}$ be a set of possible packet mutation functions, such as byte swapping and obfuscation. If the standard protocol employs packets $\{m_1, m_2, \ldots\}$, then the protocol dialect corresponding to $d_n\in \mathcal{D}$ (denoted as a dialect generating function) uses packets $\{d_n(m_1), d_n(m_2), \ldots\}$ for communication under the same communication rules. Thus, each protocol dialect is uniquely defined by a distinct dialect generating function $d_n\in \mathcal{D}$.
\end{definition}

From the given definition, protocol dialect can increase the variation of system attributes to improve communication security. We design a protocol dialect customization scheme to ensure the evolving of dialects used for every handshake. \NAME allows the client and server, who are aware of the same protocol dialect varying pattern to communicate with each other correctly. Any infiltrated malicious packets sent by an attacker will be detected and discarded by the client/server without noticing the mutation pattern.

In order to develop a moving target defense using protocol dialects, the critical challenge is keeping both client and server synchronized during communication. Under the ideal circumstance, both sides will follow the same protocol dialect variation pattern. Thus, the client/server can send and receive packets using the same dialect during one handshake. However, supposing there are disruptions in the channel (e.g., losing packets in transmission), the client and server will lose synchronization without an appropriate synchronization mechanism, leading to the disorder of selecting dialects within coming handshakes. This paper borrows the idea from the self-synchronizing stream cipher and redesigns its key mechanism in \NAME. We use previously sent packets as the reference to guarantee that the dialect selected by the client and server will rematch after a limited time of error propagation when disruptions or attacks happen.

\section{System Design}

\begin{figure}[t]
	\centering
	\includegraphics[width=0.9\textwidth]{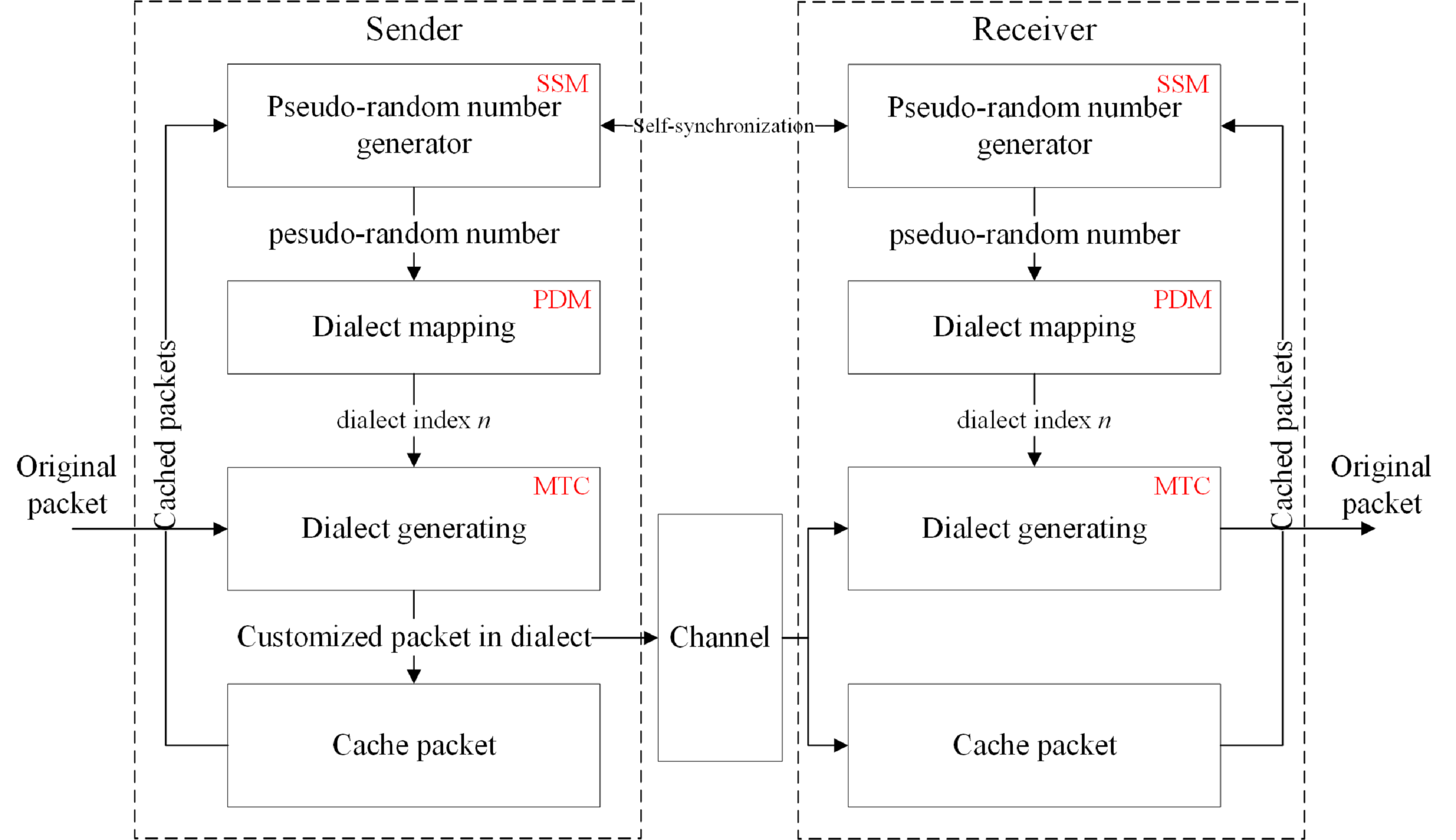}
	\caption{\NAME System Diagram.}
	\label{fig:sys}
\end{figure}

\NAME consists of three major modules: (i) Moving Target Customization (MTC), (ii) Self-Synchronization Mechanism (SSM), and (iii) Protocol Dialect Management (PDM). Its architecture is illustrated in Figure \ref{fig:sys}. When commands (e.g., GET command to retrieve a file or LS command to display a local directory in FTP) are received by a client, PDM automatically selects a protocol dialect (with a dialect index $n$) for the corresponding packets, and SSM ensures that the server and the client remain synchronized with respect to the dialect they decide to use. We note the dialect selections must be unpredictable to eves-droppers in order to prevent attacks on the protocol dialect, such as the denial of service attacks and command injection attacks. To this end, a pair of pseudo-random number generators that are self-synchronous are employed by SSM at the server and the client, respectively. Then, the consistent hashing mapping function is used in PDM to map the output of such pseudo-random number generators to proper dialect index $n$. Next, PDM instructs MTC to generate customized communication packets using the selected dialect generating function $d_n$ -- which are pre-installed on the server and the client -- and send those packets to the receiver. In particular, the sender's MTC module applies $d_n$ on each out-going packet while the receiver's MTC module employs its inverse function $d_n^{-1}$ to destruct the dialect and recover the standard packets for processing at the receiver.
\subsection{Moving Target Customization}
\label{sec:sys-mtd}

In our system, the MTC module can utilize any function $d_n$ to create protocol dialects, as long as (i) $d_n$ can sufficiently (i) mutate any communication packets $m$ and (ii) the inverse $d_n^{-1}$ exists and can be used to recover the original packets $m$, i.e.,
\begin{equation}
	\begin{array}{lr}
		m'=d_n(m) \ {\rm and} \
		m=(d_n)^{-1}(m'), \ \forall m\label{eq:customize}
	\end{array},
\end{equation}
where $m$ and $m'$ are the original and customized packets containing application layer information. To demonstrate the key ideas, we design two groups of dialects which are (i) shuffling dialect and (ii) packet-splitting dialect. 

%\subsubsection{Shuffling: }

\textbf{Shuffling:} The shuffling function aims to generate various protocol dialects by switching randomly selected segments from the original packet. Assuming a total number of $s$ switching-available segments within one packet, the mutations that we can create are the permutation of the $s$ segments. Each of these permutations is considered as one specific dialect. In practice, we often carefully select the subset of mentioned permutations to achieve better performance, as not the shuffling of every available segment will be valid given the possibility that some segments contain the same information or vital information that cannot be arbitrarily moved.

\textit{Example 1:} In this example, we illustrate byte-shuffling by defining three parameters, i.e., position, length and offset. \textit{Position} decides starting byte of first segment. \textit{Offset} is the distance between starting bytes of two segments. \textit{Length} determines the length of two segments. We used $p$, $l$ and $o$ to denote those three parameters accordingly. The dialect index $n$ for bytes-shuffling maps to the available combination of three mentioned parameters, noted as $n(p,l,o)$, which is predefined. Meanwhile, letting $m$ be a packet of $k$ bytes, we denote its $i$ byte by $m(i)$ for $i=1,2,\ldots,k$. A segment of the packet $m$ from $i^{th}$ to $j^{th}$ is denoted as $m(i,j)$. Therefore, we describe the bytes-shuffling function as following:
\begin{equation}
	\begin{split}
		d_{n(p,l,o)}(m)&=[m(1,p)||m(p+o,p+o+l)||m(p+l,p+o)||\\
		&\quad\ \  m(p,p+l)||m(p+o+l,k)]\\
		&=m',
		\label{eq:shuffle}
	\end{split}
\end{equation}
where $||$ denotes the separation of two segments within a packet. In equation, $m(p,p+l)$ and $m(p+o,p+o+l)$ are the two segments shuffled by function $d_{n(p,l,o)}$. The final result is the customized packets $m'$. In order to destruct the dialect and retrieve original packet $m$ on receiver side, we need to employ an inverse function of $(d_{(p,l,o)})^{-1}$, which is:
\begin{equation}
	\begin{split}
		(d_{n(p,l,o)})^{-1}(m')&=[m(1,p)||m(p,p+l)||m(p+l,p+o)||\\
		&\quad\ \  m(p+o,p+o+l)||m(p+o+l,k)]\\
		&=m.
		\label{eq:inverse_shuffle}
	\end{split}
\end{equation}

Since three parameters will be synchronous for both sides during communication, we can locate two segments we need to shuffle back and then recover the original packets. By using the bytes-shuffling function and its inverse function, we can automatically generate the desired number of dialects for packets. 

%\subsubsection{Packet-splitting: }

\textbf{Packet-splitting:} The second type of dialect is packet-splitting dialect. This will generate protocol dialects by splitting a single packet into several sub-packets of any length (smaller than the original packet). In each sub-packet, the lower-layer header remains, but it only carries part of the application layer information of the original packet. Considering that each sub-packet will receive its corresponding response from the receiver, it breaks one handshake into several handshakes. Therefore, the packet-splitting dialect will change the original handshake pattern into a customized one with multiple handshakes.

\textit{Example 2:} In this example, we split one single packet into four sub-packets. We introduce three predefined parameters $t_1$, $t_2$ and $t_3$ to denote the length of first three sub-packets correspondingly. Since the original packet has the fixed message length $k$, the length of the last sub-packets is equal to $k-t_1-t_2-t_3$. The combination of every four sub-packets is a new dialect. Therefore, we describe the packet-splitting function as following:

\begin{equation}
	\begin{split}
		d_{n(t_1,t_2,t_3)}(m)&=d_{n(t_1,t_2,t_3)}([m(1,t_1) || m(t_1, t_1+t_2) ||\\   
		&\quad\  m(t_1+t_2,t_1+t_2+t_3) || m(t_1+t_2+t_3,k)])\\
		&=[m'_1 || m'_2 || ... || m'_4],
		\label{eq:split}
	\end{split}
\end{equation}
where $m'_i$ denotes each sub-packet created by function $d_{n(t_1,t_2,t_3)}$. Each sub-packet accordingly contains a segment of the information from original packet. In order to retrieve the original packet $m$ from each sub-packet, receiver need to leverage inverse function $(d_{n(t_1,t_2,t_3)})^{-1}$ to merge all received sub-packets, which is:
\begin{equation}
	\begin{split}
		(d_{n(t_1,t_2,t_3)})^{-1}([m'_1 || m'_2 || ... || m'_4])&=(d_{n(t_1,t_2,t_3)})^{-1}([m(1,t_1)||m(t_1, t_1+t_2)||\\
		&\quad\  m(t_1+t_2,t_1+t_2+t_3)||m(t_1+t_2+t_3,k)])\\
		&=m.
		\label{eq:inverse_split}
	\end{split}
\end{equation}

Each index indicates one specific combination of three defined parameters $t_1$, $t_2$ and $t_3$. With different dialect index $n$, the length of the sub-packets will be different. Due to the synchronization mechanism of protocol dialect, both sides will synchronize and generate the same dialect index $n$ during the communication. Hence, we can collect all the sub-packets and merge them to recover original packets.

\subsection{Self-Synchronization Mechanism}
\label{sec:sys-syn}

Dialect selection by the server and client must be unpredictable to any eves-droppers and yet remain synchronized to ensure the proper functioning of the communication protocol. To this end, we propose a mechanism that leverages a pseudo-random number generator to create randomness for selecting random dialects. Similar to the idea of self-synchronizing stream ciphers, we make future dialect indexes depend on past packet values to allow the server and client to self-synchronize even under packet erasures/modifications.

We cache one past packet (denoted by $m_{i-1}$) in a local buffer at the server and the client, i.e.,
\begin{equation}
	M_i=m_{i-1}.\label{eq:cat}
\end{equation}

The Keyed-hash function is widely used in the keyed-hash message authentication code (HMAC). However, we note that rather than using the keyed-hash value for message authentication, we consider it as a pseudo-random number that can be securely generated by the client and server and compute the current dialect index from the keyed-hash value. Therefore, next, we leverage the keyed-hash function to produce a pseudo-random number based on a shared secret key $K$ known to both the server and client. For a current cached message $M_i$ and shared secret $K$, the keyed-hash function will generate a pseudo-random number $S$ from $M_i$. 

More precisely, we present the keyed-hash function we used in our implementation as following:

\begin{equation}
	\begin{split}
		S&=H((K \oplus opad) || H(K \oplus ipad) || M_i)\\
		&=H((K \oplus opad) || H(K \oplus ipad) || m_{i-1}),\label{eq:newHMAC}
	\end{split}
\end{equation}
where $H$ is cryptographic hash function such as MD5, and $\oplus$ denotes bitwise exclusive or (XOR). According to RFC 2104, outer padding $opad$ consists of repeated bytes valued 0x5c, and inner padding $ipad$ consists of repeated bytes valued 0x36. As shown in equation \ref{eq:newHMAC}, the input becomes shared secret key $K$ and cached past packet $M_i$, and those work as random seed for each iteration.

\begin{figure}[t]
	\centering
	\includegraphics[width=0.75\textwidth]{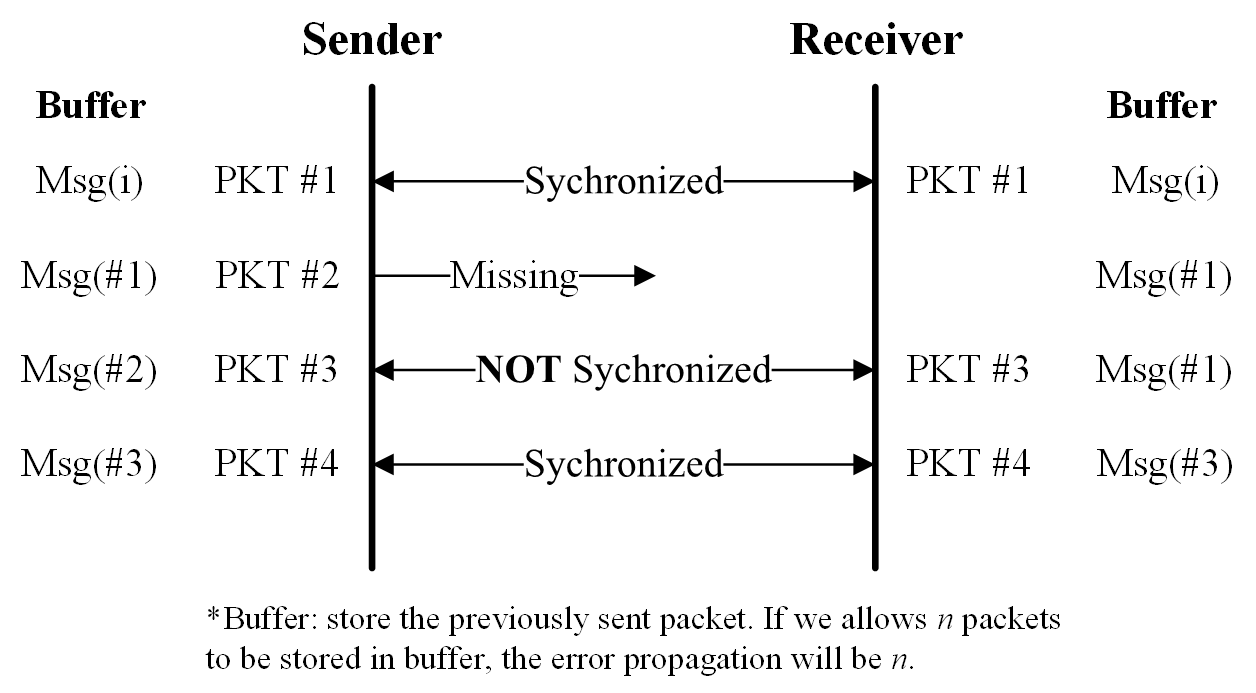}
	\caption{An illustration of the proposed self-synchronization mechanism.}
	\label{fig:syn}
\end{figure}

The modified keyed-hash function used in this paper is just one among many possible candidates from the available pseudo-random functions family. Due to the security properties provided by keyed-hash functions, the resulting pseudo-random number $S$ is unpredictable to any eves-dropper who does not have access to the shared secret key $K$. It is also apparent that the server and the client will eventually resume synchronization even under packet erasures and modification. As illustrated in Figure~\ref{fig:syn}, as long as the past packets stored in local buffers are the same for both client and server, their pseudo-random number generator would compute the same pseudo-random number. Now the self-synchronous property is realized between client and server and can always be achieved if making the final dialect index depend on past packet values. 

We note that similar to self-synchronizing stream ciphers, caching more than one previous packet and concatenate them as new $M_i$ will leverage the contents of previously sent packets, which increases the complexity of modified keyed-hash function (i.e., equation \ref{eq:newHMAC}) and efficiency against brute force attacks aiming at the self-synchronizing mechanism. In contrast, it takes longer to re-synchronize the server and the client under network errors.
\subsection{Protocol Dialect Management}
\label{sec:sys-manage}

\begin{figure}[t]
	\centering
	\includegraphics[width=0.5\textwidth]{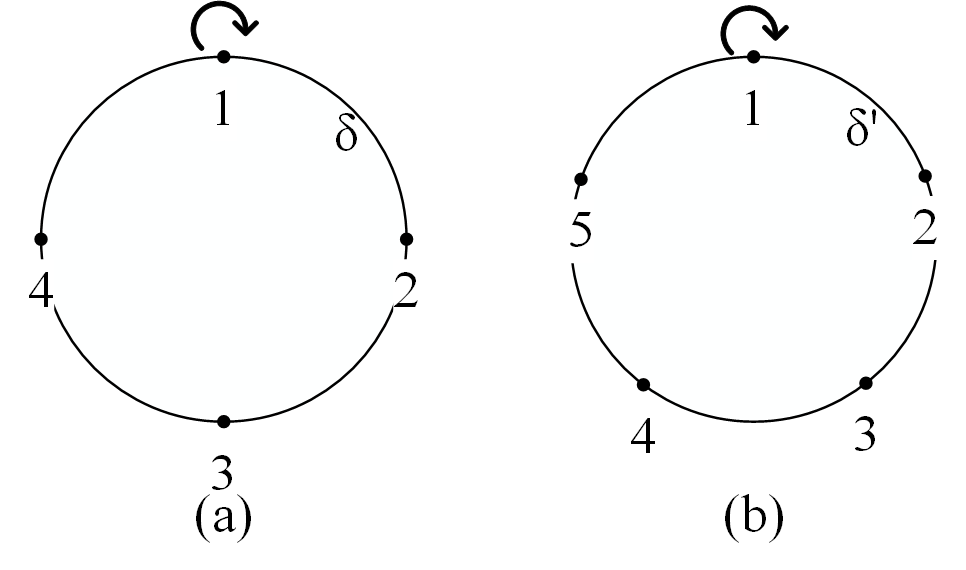}
	\caption{Consistent hashing mapping before (a) and after (b) adding new dialect.}
	\label{fig:map}
\end{figure}

Finally, we need to map the pseudo-random number got from equation \ref{eq:newHMAC} to dialect index $n$. We implement the idea borrowed from consistent hashing in keyspace partitioning. Keyspace partitioning originally aims at mapping keys to nodes in a distributed hash table. We keep the concept but replace the keys with the pseudo-random number we had and nodes with protocol dialect indexes. Therefore, we design consistent hashing mapping that treats protocol dialect indexes as points on a circle, and $\delta(n, n+1)$ is the distance between two points, which is traveling clockwise around the circle from index $n$ to $n+1$, as shown in Figure \ref{fig:map}. Assuming the pseudo-random number value is $S$ and distance $\delta$ between every two indexes are equally distributed, in this case, consistent hashing mapping is presented as following:
\begin{equation}
	n=S//(S_{max}//n_{max}+1)+1,\label{eq:mapping}
\end{equation}
where $S_{max}$ is the maximum pseudo-random number value we can generate, and $//$ denotes floor division. Considering that the total number of protocol dialects is $n_{max}$, we can decide the certain dialect with index $n$ that should be applied to the current packet by computing equation \ref{eq:mapping}.

Due to the different protocol dialects that we created, consistent hashing mapping gives us the flexibility to add new dialects or drop existing dialects. Without making significant changes in the mapping function, we can minimize the update of equation \ref{eq:mapping} and thereby increase the efficiency of the design. 

\textit{Example 3:} Given the example illustrated in Figure \ref{fig:map}, the circle in this figure represents the range of pseudo-random numbers ranging from 0 to $S_{max}$. We have four existed dialects in use shown in Figure \ref{fig:map} (a), and each of them takes a quarter of the circle. Assuming that we add a new dialect 5 as a candidate, in equation \ref{eq:mapping}, the only changed parameter is the total number of protocol dialects $n_{max}$, which should increase from 4 to 5. Representing in Figure \ref{fig:map} (b), the distance $\delta$ will decrease to $\delta'$ which is one-fifth of the circle. Therefore, we can update $n_{max}$ value in the equation without modifying any parameters else and thereby mitigate the impact made by adding/dropping of dialect in our system.

\subsection{Security Analysis}
\label{sec:sys-anal}

IoT protocol MQTT is widely applied in many popular IoT applications, such as IBM Watson IoT Platform and AWS IoT services. Four phases are in an MQTT session: (i) connection, (ii) authentication, (iii) communication, and (iv) termination. After establishing the connection by creating a TCP/IP connection with the server (broker) on a predefined port, the client will send the \textit{CONNECT} packet that includes user identification to start the MQTT communication. In this step, an attacker can launch a Denial of Service (DoS) attack by sending many \textit{CONNECT} requests continuously and thereby maliciously making the server busy as in requests flooding, which impairs the availability of the system~\cite{firdous2017modelling}. As the server is not able to differentiate the normal \textit{CONNECT} and the malicious \textit{CONNECT} packets, on receiving the flooding request messages, the broker starts to acknowledge all with \textit{CONNACK} message. Assuming many connection requests that arrive simultaneously, the server's buffer will be drained out, and the server will not be capable of processing new incoming requests. In our design, we will apply the \NAME on \textit{CONNECT} packets. Before the server receiving incoming \textit{CONNECT} packets from the client, the system will check whether the dialect index of the current packet matches that of the server. For malicious \textit{CONNECT} packets, no dialect will be applied, and therefore they will fail the dialect index check. In this case, the server will refuse to start MQTT communication when receiving malicious request messages, saving its resources for processing valid requests. 

Attacks such as command injection attacks and cipher suite downgrade attacks can intentionally erase or alter the communication packets on a public channel, damaging the integrity of communication. In the light-weight controller area network (CAN) protocol, a message-based IoT protocol designed to allow microcontrollers and devices to communicate with each other's applications, no encryption or authentication is initially applied due to their high cost. Thus, it is vulnerable and threatens by safety issues such as command injection attack~\cite{liu2017vehicle}. In this type of attack, attackers will use reprogrammed electronic control units (ECU) to send malicious packets to the CAN bus. These malicious packets are fabricated and injected with forged ID and data to distract victim ECUs or make them execute malformed actions. For some other protocols, a standard mechanism such as SSL/TLS can be used to protect communication security. However, aiming to vulnerability existed in SSL/TLS, man in the middle (MITM) attackers can launch cipher suite downgrade attacks by abandoning the \textit{ClientHello} packet sent by the genuine client and replacing it with a malformed packet containing lower-version TLS~\cite{lee2020return,sjoholmsierchio2020strengthening}. In consequence, the server will start the communication with the client using insecure lower-version TLS. We propose the \NAME that will guarantee the evolution of packet dialect patterns during communication. Even though an attacker leverage the fetched information to launch an attack, malicious packets such as command injection packet or lower-version \textit{ClientHello} packet will not be processed by the target because they cannot match the dialect evolution pattern of the receiver without acquiring knowledge of \NAME. 

In the above threat models, if packets are missing in an unstable communication channel or maliciously erased/altered by an attacker, it will affect the current packet in communication and drive the moving target defense of the client and server into an asynchronous state. The self-synchronization mechanism in our proposed solution guarantees the resilience of our system under packet erasures and modifications. Supposing there is a packet missing or modified, according to Equation (\ref{eq:newHMAC}), it will only affect the dialect selection of the next packet (i.e., causing the server and the client to select different dialects for the next packet). Thus, such an error only propagates once. The server and the client will resume synchronization after a short, finite, and predictable transient time (when the erased or corrupted packet moves out of the local buffer). 

As shown in Figure \ref{fig:syn}, supposing the packet \textit{PKT \#2} is missing/erased, when transferring the \textit{PKT \#3}, previously cached packets stored in the local buffer of both sides are different. The sender assumes \textit{PKT \#2} was successfully sent and updates it into its local buffer, while the cached packets in the receiver's local buffer remain unchanged. The total number of past packets we cached is one, and therefore the error will propagate for only one handshake from \textit{PKT \#2} to \textit{PKT \#3}. After that, the sender and receiver return to a synchronous state again as the cached error packets move out of their local buffers. We note that if expanding the total number of cached packets in the local buffer to $h$, the error will propagate for $h$ handshakes accordingly.

\section{Implementation}

We implement a prototype of \NAME on FTP and MQTT, which includes the following main components. 

{\em Dialect Customization and Management:} We implement our dialect customization functions (e.g., shuffling-byte) in FTP and MQTT. Its input parameters are predefined and bundled together. We use indexes to indicate those parameter combinations. The index is decided by the consistent hashing mapping function in the PDM module. It determines which dialect is used for each communication packet.

{\em Pseudo-Random Number Generator:} We implement a keyed-hash function as the pseudo-random number generator. The output pseudo-random number will be fed into the consistent hashing mapping function to compute the corresponding dialect index. 

{\em Self-Synchronization Module:} We implement our SSM module and cache one previous packet, which implies that any network errors will propagate for only one cycle during communication. The buffer will be filled with one predefined initial packet at the beginning of the program execution.

\section{Evaluation}
\label{sec:eval}

\begin{table*}[t]
	\caption{Examples of byte-shuffling dialect evolution pattern, including dialect index and parameters for the dialect generating function, computed for different FTP GET commands.}
	\label{tab:syn_ftp}
	\centering
	\begin{threeparttable}
		\begin{tabular}{c|c|c|ccc}
			\Xhline{1pt}
			Original packet&Customized packet&Index&Position&Length&Offset\\[1pt]
			\Xhline{.5pt}
			rget,sample.txt&r,etgsample.txt&8&1&1&3\\[1pt]
			rget,sample.txt&tger,sample.txt&7&0&1&3\\[1pt] 
			rget,blog.css&etrg,blog.css&4&0&2&2\\[1pt]
			rget,template.pdf&etrg,template.pdf&4&0&2&2\\[1pt]
			\Xhline{1pt}
		\end{tabular}
		%		\begin{tablenotes}
		%\footnotesize
		%\item[1] Pre-defined mapping divisors are listed accordingly: x=4; y=3; z=4.
		%\item[2] Total possible dialects: xyz=48.
		%		\end{tablenotes}
	\end{threeparttable}
\end{table*}

\begin{table*}[t]
	\caption{Examples of packet-splitting dialect evolution pattern, including dialect index and parameters for the dialect generating function, computed for different FTP GET commands.}
	\label{tab:syn_ftp2}
	\centering
	\begin{threeparttable}
		\begin{tabular}{c|c|c|cccc}
			\Xhline{1pt}
			\multirow{2}*{Original packet}&\multirow{2}*{Customized sub-packets}&\multirow{2}*{Index}&\multicolumn{4}{c}{Length}\\
			&&&Pkt1&Pkt2&Pkt3&Pkt4\\[1pt]
			\Xhline{.5pt}
			rget,sample.txt&[r][ge][t,][sample.txt]&6&1&2&2&10\\[1pt]
			rget,sample.txt&[r][ge][t][,sample.txt]&2&1&2&1&11\\[1pt] 
			rget,blog.css&[r][ge][t][,blog.css]&2&1&2&1&9\\[1pt]
			rget,template.pdf&[rg][et][,][template.pdf]&4&2&2&1&12\\[1pt]
			\Xhline{1pt}
		\end{tabular}
	\end{threeparttable}
\end{table*}

In this section, we evaluate the performance of \NAME and the effectiveness of the proposed moving target defense.

\textbf{Experiment Setup:} Our experiments are conducted on a 2.60GHz Intel(R) Core(TM) CPU i7-9750H machine with 16 Gigabytes of memory. The operating system is Ubuntu 18.04 LTS. We perform MTC, SSM, and PTD modules on the selected communication protocol.

\begin{figure}[t]
	\centering
	\includegraphics[scale=0.5]{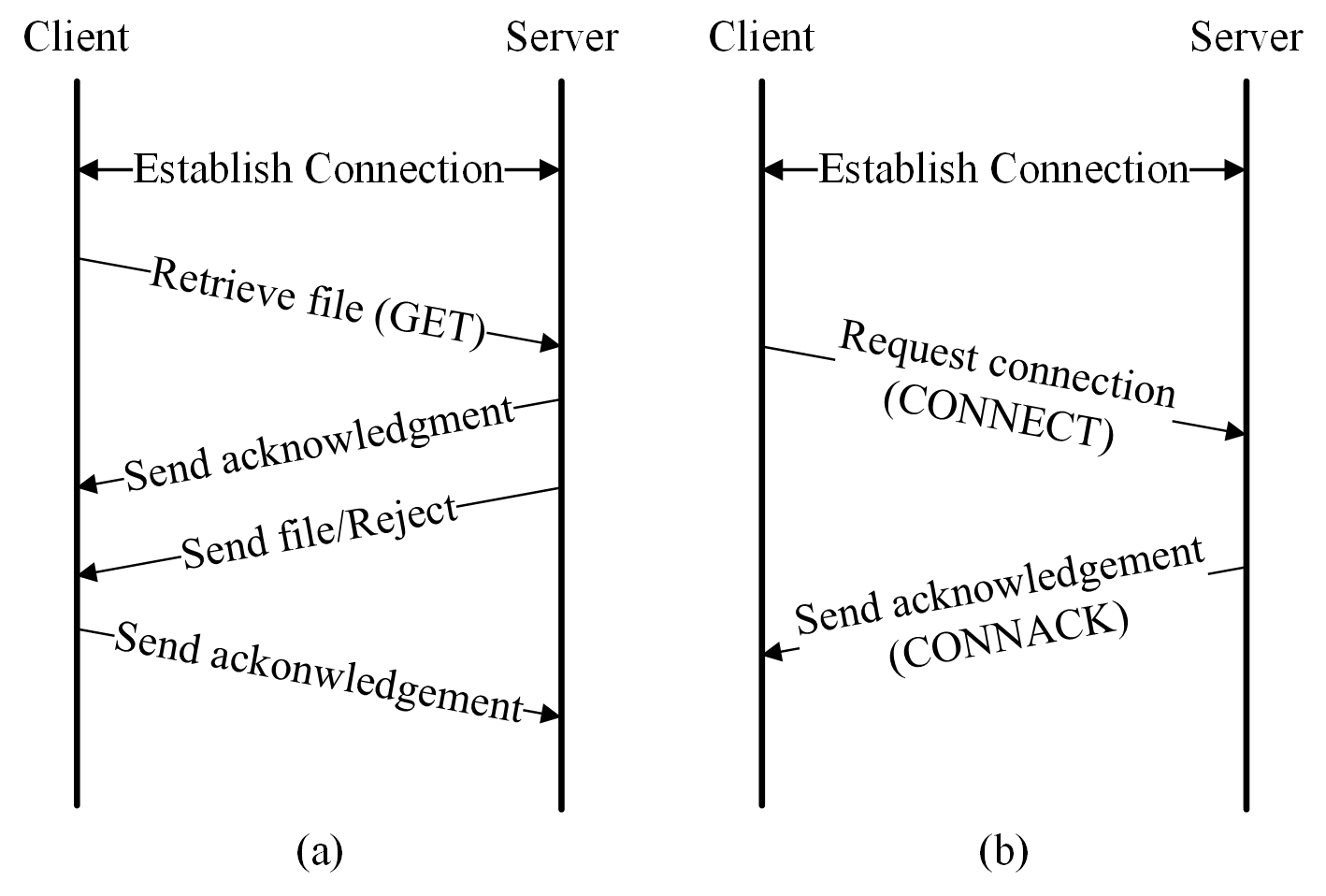}
	\caption{Timing diagram of GET command in FTP (a) and CONNECT action in MQTT (b).}
	\label{fig:timing}
\end{figure}

\begin{figure}[t]
	\centering
	\includegraphics[width=0.5\textwidth]{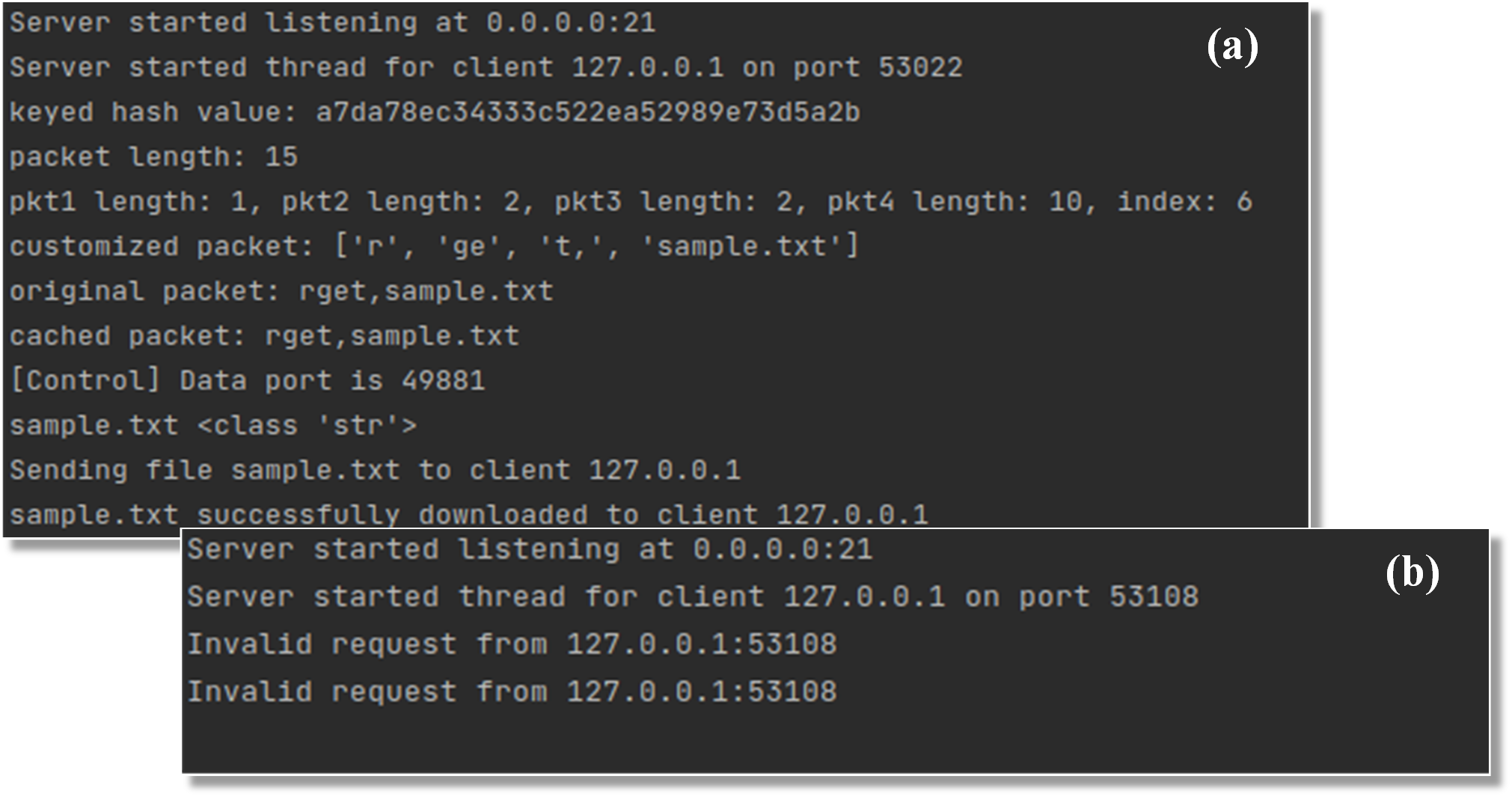}
	\caption{Results of establishing connection with server by genuine FTP client (a) and attacker FTP client (b).}
	\label{fig:cmp}
\end{figure}

\textbf{Target Protocol: FTP } FTP is a standard network protocol used for the transfer of computer files between the client and the server. Many FTP clients and automated utilities have been developed for desktops, servers, mobile devices, and hardware. As a target protocol for our proof of concept evaluation, FTP offers two main benefits: (i) It is a lightweight protocol having better performance and easier to test, and (ii) it does not have many complex features, making it easier to customize and analyze.

FTP packet format contains IP header, TCP header, and FTP message contents. We programmed a standard FTP client and server and applied our implementation to them. For instance, when a GET command is received on the server, the corresponding file (if file exists) or rejection message (if the file does not exist) will be sent to the client via four handshakes, as shown in Figure~\ref{fig:timing} (a). After adding the MTC module on both server and client, the packet for transferring the GET command will be cast into a proper dialect during each handshake.

Table~\ref{tab:syn_ftp} and \ref{tab:syn_ftp2} show byte-shuffling dialect and packet-splitting dialect as well as the evolution process of key parameters during four consecutive file transfers. Due to the different pseudo-random numbers being generated for each message/packet, the PDM module will map the number into different dialect indexes, resulting in different dialect selections and different input parameters of the MTC module for customizing packets. On the server-side, after performing the inverse process, the original packets are recovered, and the server will send the requested file to the client, as shown in Figure~\ref{fig:cmp} (a).

To demonstrate the effectiveness of our moving target defense, we create an attacker FTP client that implements a spoofed (and fixed) dialect of the FTP. The attacker client connects to the FTP server equipped with our moving target defense and starts fabricating and sending the malicious command to the server to launch the command injection attack. The result are presented in Figure~\ref{fig:cmp} (b). It is obvious that the command injection attack fails to work. Since the attacker's client protocol dialect pattern is fixed, it is unable to understand the dialect evolution pattern generated in our moving target defense applied on the GET command. We prove that our method can effectively defend the system from such attacks.

\begin{figure}[t]
	\centering
	\includegraphics[width=0.56\textwidth]{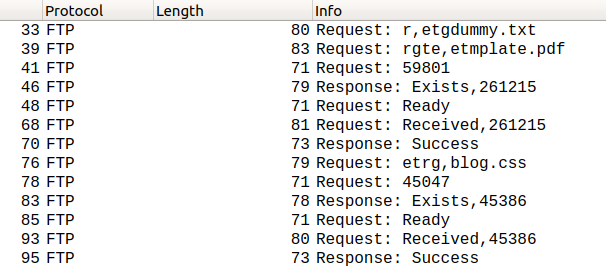}
	\caption{Packets captured by Wireshark showing error propagation equal to the total number of previously sent packets cached in the buffer.}
	\label{fig:missing}
\end{figure}

\begin{table}[t]
	\caption{Self-synchronization mechanism demonstration with missing Pkt \#1.}
	\label{tab:syn_missing}
	\begin{threeparttable}
		\begin{tabular}{c|cc|cc|c}
			\Xhline{1pt}
			Iteration&Client&Cached Pkt&Server&Cached Pkt&Status\\[1pt]
			\Xhline{.5pt}
			1&Pkt \#1&[Pkt \#p]&|&[Pkt \#p]&Synchronized, Nothing received\\[1pt]
			2&Pkt \#2&[Pkt \#1]&Pkt \#2&[Pkt \#p]&Not synchronized\\[1pt]
			3&Pkt \#3&[Pkt \#2]&Pkt \#3&[Pkt \#2]&Synchronized\\[1pt]
			4&Pkt \#4&[Pkt \#3]&Pkt \#4&[Pkt \#3]&Synchronized\\[1pt]
			\Xhline{1pt}
		\end{tabular}
		\begin{tablenotes}
			\footnotesize
			\item[1] We assume \textit{Pkt \#1} is erased/missing during communication.
			\item[2] \textit{Pkt \#p} is previously sent packet that has already been cached in local buffer.
		\end{tablenotes}
	\end{threeparttable}
\end{table}

Next, we evaluate the effectiveness of our self-synchronization mechanism under erased/missing packets. In particular, we choose to store only one past packet in the buffer for self-synchronization. As is shown in Figure~\ref{fig:missing}, suppose that, due to some network error or intentional attack, one GET command packet before frame \#33 gets lost during communication (Thus, this missing packet cannot be captured by Wireshark in the screenshot). As we can observe, for the second GET command packets that the client sent, which in Figure \ref{fig:missing} corresponds to frame \#33, the server generates no response. Till now, the dialect selection at the client and the server are not synchronized because the previously sent packet cached in the local buffer is different, resulting in selecting different dialect for \textit{Pkt \#2} as shown in Table~\ref{tab:syn_missing} Row 3. However, starting from the third packet, which is frame \#39 in Figure \ref{fig:missing} and Row 4 in Table \ref{tab:syn_missing}, both sides are successfully re-synchronized, as the different packets move out of their local buffers after one (network) cycles, and as a result, the same previously sent packets are again shared between them. It demonstrates \NAME's ability to self-synchronize under packet erasures or in an unstable communication channel.

Finally, we evaluate the overhead of \NAME, by comparing it with the execution overhead of a standard FTP implementation on the server-side, concerning the network, CPU, and memory overhead. To this end, we use the \textit{time} command in the terminal to monitor several performance indexes such as running time and maximum resident set size. To avoid potential statistical bias, we run each experiment four times and record the average overhead in Table~\ref{tab:overhead}. For each test, we write a script to execute one billion FTP commands randomly (e.g., using GET to continuously retrieve a small text file from the server to the client).

\begin{table}[t]
	\caption{Evaluating the overhead of designed moving target defense in FTP.}
	\label{tab:overhead}
	\centering
	\begin{threeparttable}
		\begin{tabular}{c|cc}
			\Xhline{1pt}
			Performance Index&Original FTP&Modified FTP\\[1pt]
			\Xhline{.5pt}
			System time/sec&0.57&0.60\\[1pt]
			Elapsed time/sec&42.64&44.53\\[1pt]
			Percent of CPU this job got&1\%&1\%\\[1pt]
			Maximum resident set size/KB&6402&7683\\[1pt]
			\Xhline{1pt}
		\end{tabular}
		\begin{tablenotes}
			\footnotesize
			\item[1] System time: time spent in kernel mode while running the program.
			\item[2] Maximum resident set size: Maximum memory space occupied while running the program.
		\end{tablenotes}
	\end{threeparttable}
\end{table}

As it is shown in the Table~\ref{tab:overhead}, while a number of functions (including dialect generating function, pseudo-random function, and consistent hashing mapping function) are added in FTP to support our moving target defense, the overhead is almost negligible as compared to a standard FTP program. More precisely, the execution overhead (as measured by elapsed time) increases by 4.43\%, while the maximum resident set size we need compared to standard FTP implementation increase by 1281 KB.

\textbf{Target Protocol: MQTT } MQTT is a standard lightweight IoT protocol for transporting messages among IoT devices. Due to its small size, it is designed to provide efficient message delivery for the network where the bandwidth is limited. Apart from the similar reasons we mentioned for choosing FTP, we select MQTT as another target protocol for analysis because it is the typical IoT protocol used by many IoT devices, and shares many common characters compared with other IoT protocols.

\begin{figure}[t]
	\centering
	\includegraphics[width=0.5\textwidth]{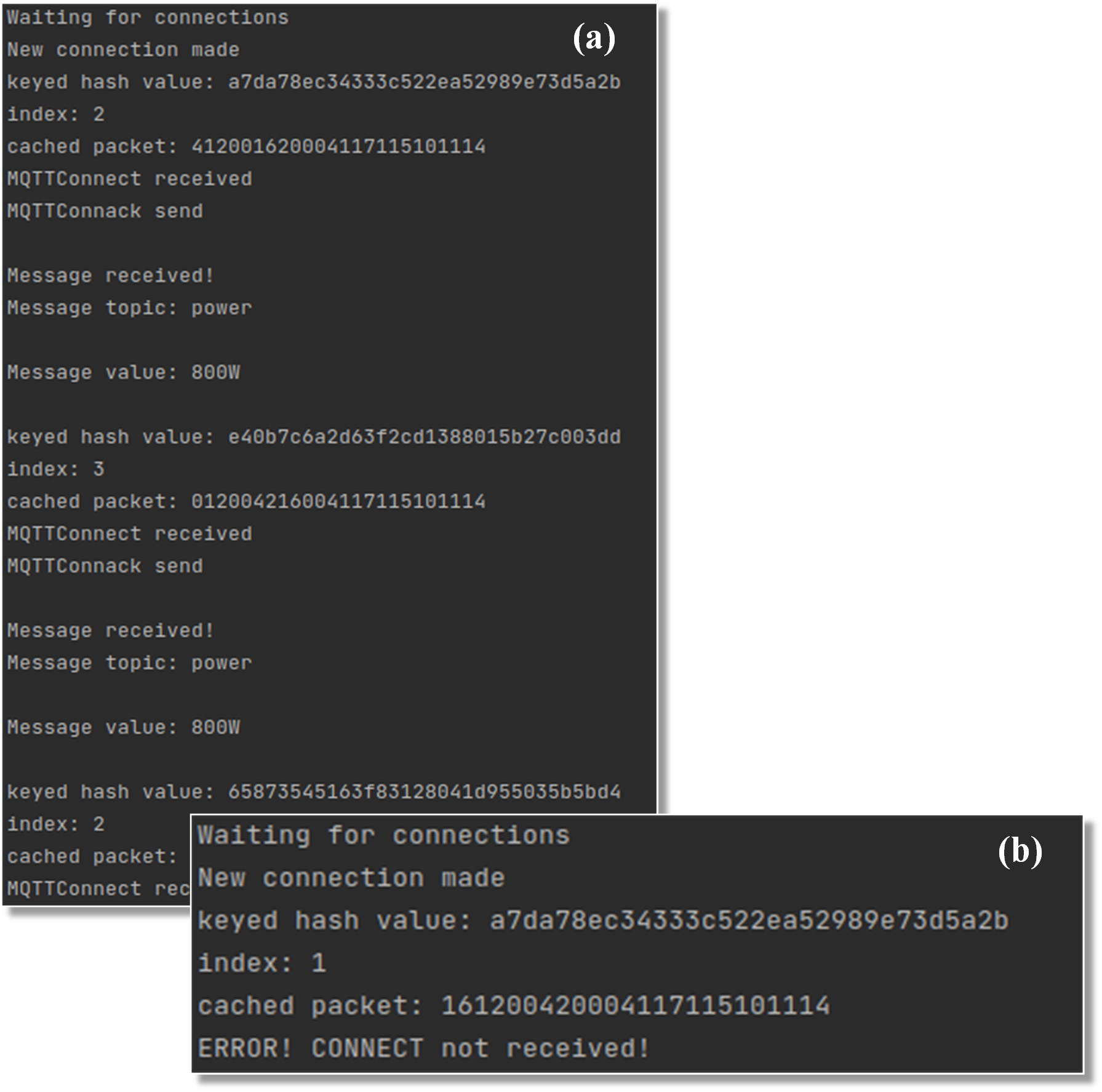}
	\caption{Results of establishing connection with server (broker) by genuine MQTT client (a) and attacker MQTT client (b).}
	\label{fig:cmp_mqtt}
\end{figure}

MQTT usually runs over TCP/IP. Besides IP header and TCP header, an MQTT packet also contains a fixed header (including control field, e.g., \textit{CONNECT}/\textit{CONNACK}, and packet length field), variable length header, and payload on the application layer. We programmed a standard MQTT client and server (broker) and applied \NAME on them. The timing diagram of establishing the MQTT connection is shown in Figure~\ref{fig:timing} (b).  The client will send the \textit{CONNECT} to the server to request MQTT communication. After receiving \textit{CONNECT} packet, server will response with \textit{CONNACK} to confirm the connection. We add the MTC module using the byte-shuffling function on both sides, and thus the \textit{CONNECT}/\textit{CONNACK} packet will be cast into an appropriate dialect during each handshake.

To demonstrate the capability of the modified MQTT system in defending the DoS attack, we implement an attacker MQTT client that tries to flush the modified server by continuously sending \textit{CONNECT} packets with different identities. The MTC module uses the byte-shuffling function, which will shuffle the control field with other fields in the \textit{CONNECT} packet. As shown in Figure~\ref{fig:cmp_mqtt} (b), the attacker fails to use multiple malicious \textit{CONNECT} packets to exhaust the server's buffer space because these packets will fail the dialect index check before getting accepted. In contrast, as shown in Figure~\ref{fig:cmp_mqtt} (a), the genuine client is able to successfully start communication and publish information on the server due to their having the synchronized dialect evolution pattern. Therefore, we can apply \NAME on IoT protocol MQTT to prevent the system from some types of DoS attacks.

\section{Related Work}

\textbf{Pseudo-random number generator:} Pseudo-random number generator (PRNG) is an algorithm for generating a sequence of numbers that have properties approximating the properties of random-number sequences. Since we need to implement \NAME with randomly varying moving targets, the basic idea is using PRNG to output a random number for self-synchronization and dialect selection. Many previous works~\cite{salmon2011parallel,akhshani2014pseudo,wang2019pseudo,lee2001elliptic} provide us with insight of choosing PRNG, such as PRNG based on logistic chaotic system~\cite{wang2019pseudo} and elliptic curve PRNG~\cite{lee2001elliptic}.

\textbf{Self-synchronizing stream cipher:} Stream cipher~\cite{hell2007grain,paul2011rc4,mannai2018new} is a symmetric key cipher where the digits of plain text are combined with a pseudo-random cipher digit stream (which is the keystream), including two types: (i) synchronous stream cipher (SSC), and (ii) self-synchronizing stream cipher (SSSC). The latter uses previous ciphertext digits to compute the keystream. In paper~\cite{maurer1991new}, several alternative design approaches for SSSCs are proposed that are superior to the design based on the block cipher with respect to encryption speed and security. Joan Daemen et al. introduce another design approach for hardware-oriented SSSCs named Moustique~\cite{daemen2008self}. Besides, many other discussions are about SSSCs, which can be found in papers~\cite{khazaei2008new,millerioux2010self}. 

\textbf{Moving target defense (MTD):} Plenty of previous works about MTD have different areas of focus respectively. Some of them are about leveraging lower-level system configurations, which provide insights into our design. In OpenFlow random host mutation~\cite{jafarian2012openflow}, Jafarian et al. provided an MTD architecture that transparently mutates IP addresses with randomness. RPAH~\cite{luo2015rpah} achieved MTD by constantly changing IP addresses and ports to realize random port and address hoping. On protocol level, Ghost-MTD~\cite{park2020ghost} applied mutation on protocols to achieve MTD, while the protocol mutation pattern should be pre-defined and pre-shared between client and server. 

\textbf{Security risks of IoT protocols:} Internet of things is growing rapidly and reaches a multitude of different domains such as environmental monitoring, smart home, and automatic driving. However, many low-end IoT products do not usually have strong security mechanisms embedded~\cite{firdous2017modelling,hartzell2017automobile}. Hence, threats such as leakage of sensible information, DoS attacks, and unauthorized network access attacks bring a severe safety issue to the IoT system~\cite{meneghello2019iot}. Some researches are conducted previously to mitigate the threat of attacks targeting IoT protocols and devices. Besides, in the survey~\cite{nebbione2020security}, Nebbione et al. provided some basic ideas for solving the security issues in many IoT protocols. 

\section{Conclusion}

We design and evaluate a novel moving target defense framework, MPD, which aims to generate customized dialects during communication and dynamically select different dialects in a self-synchronous manner. Our experiment results using FTP and MQTT indicate that \NAME is able to harden the security while incurring low execution overhead effectively.

%For future work, we plan to integrate \NAME into other communication/IoT protocols and explore different pseudo-random number generators.

%\vspace{0.5in}
\newpage

\renewcommand{\bibsection}{\section*{References}}
\bibliographystyle{splncsnat}
\bibliography{ref}

\begin{thebibliography}{35}
\providecommand{\natexlab}[1]{#1}
\providecommand{\url}[1]{\texttt{#1}}
\providecommand{\urlprefix}{}

\bibitem[{Akhshani et~al.(2014)Akhshani, Akhavan, Mobaraki, Lim, and
  Hassan}]{akhshani2014pseudo}
Akhshani, A., Akhavan, A., Mobaraki, A., Lim, S.C., Hassan, Z.: Pseudo random
  number generator based on quantum chaotic map.
\newblock Communications in Nonlinear Science and Numerical Simulation 19(1),
  101--111 (2014)

\bibitem[{Al-Shaer(2011)}]{al2011toward}
Al-Shaer, E.: Toward network configuration randomization for moving target
  defense.
\newblock In: Moving Target Defense, pp. 153--159. Springer (2011)

\bibitem[{Andy et~al.(2017)Andy, Rahardjo, and Hanindhito}]{andy2017attack}
Andy, S., Rahardjo, B., Hanindhito, B.: Attack scenarios and security analysis
  of mqtt communication protocol in iot system.
\newblock In: 2017 4th International Conference on Electrical Engineering,
  Computer Science and Informatics (EECSI). pp. 1--6. IEEE (2017)

\bibitem[{Barrantes et~al.(2003)Barrantes, Ackley, Forrest, Palmer, Stefanovic,
  and Zovi}]{barrantes2003randomized}
Barrantes, E.G., Ackley, D.H., Forrest, S., Palmer, T.S., Stefanovic, D., Zovi,
  D.D.: Randomized instruction set emulation to disrupt binary code injection
  attacks.
\newblock In: Proceedings of the 10th ACM conference on Computer and
  communications security. pp. 281--289 (2003)

\bibitem[{Cho et~al.(2020)Cho, Sharma, Alavizadeh, Yoon, Ben-Asher, Moore, Kim,
  Lim, and Nelson}]{cho2020toward}
Cho, J.H., Sharma, D.P., Alavizadeh, H., Yoon, S., Ben-Asher, N., Moore, T.J.,
  Kim, D.S., Lim, H., Nelson, F.F.: Toward proactive, adaptive defense: A
  survey on moving target defense.
\newblock IEEE Communications Surveys \& Tutorials 22(1), 709--745 (2020)

\bibitem[{Daemen and Kitsos(2008)}]{daemen2008self}
Daemen, J., Kitsos, P.: The self-synchronizing stream cipher moustique.
\newblock In: New stream cipher designs, pp. 210--223. Springer (2008)

\bibitem[{Firdous et~al.(2017)Firdous, Baig, Valli, and
  Ibrahim}]{firdous2017modelling}
Firdous, S.N., Baig, Z., Valli, C., Ibrahim, A.: Modelling and evaluation of
  malicious attacks against the iot mqtt protocol.
\newblock In: 2017 IEEE International Conference on Internet of Things
  (iThings) and IEEE Green Computing and Communications (GreenCom) and IEEE
  Cyber, Physical and Social Computing (CPSCom) and IEEE Smart Data
  (SmartData). pp. 748--755. IEEE (2017)

\bibitem[{Haripriya and Kulothungan(2019)}]{haripriya2019secure}
Haripriya, A., Kulothungan, K.: Secure-mqtt: an efficient fuzzy logic-based
  approach to detect dos attack in mqtt protocol for internet of things.
\newblock EURASIP Journal on Wireless Communications and Networking 2019(1), 90
  (2019)

\bibitem[{Hartzell and Stubel(2017)}]{hartzell2017automobile}
Hartzell, S., Stubel, C.: Automobile can bus network security and
  vulnerabilities.
\newblock Univ. Washington, Seattle, WA, USA, Tech. Rep  (2017)

\bibitem[{Hell et~al.(2007)Hell, Johansson, and Meier}]{hell2007grain}
Hell, M., Johansson, T., Meier, W.: Grain: a stream cipher for constrained
  environments.
\newblock International journal of wireless and mobile computing 2(1), 86--93
  (2007)

\bibitem[{Hund et~al.(2013)Hund, Willems, and Holz}]{hund2013practical}
Hund, R., Willems, C., Holz, T.: Practical timing side channel attacks against
  kernel space aslr.
\newblock In: 2013 IEEE Symposium on Security and Privacy. pp. 191--205. IEEE
  (2013)

\bibitem[{Jafarian et~al.(2012)Jafarian, Al-Shaer, and
  Duan}]{jafarian2012openflow}
Jafarian, J.H., Al-Shaer, E., Duan, Q.: Openflow random host mutation:
  transparent moving target defense using software defined networking.
\newblock In: Proceedings of the first workshop on Hot topics in software
  defined networks. pp. 127--132 (2012)

\bibitem[{Kc et~al.(2003)Kc, Keromytis, and Prevelakis}]{kc2003countering}
Kc, G.S., Keromytis, A.D., Prevelakis, V.: Countering code-injection attacks
  with instruction-set randomization.
\newblock In: Proceedings of the 10th ACM conference on Computer and
  communications security. pp. 272--280 (2003)

\bibitem[{Khazaei and Meier(2008)}]{khazaei2008new}
Khazaei, S., Meier, W.: New directions in cryptanalysis of self-synchronizing
  stream ciphers.
\newblock In: International Conference on Cryptology in India. pp. 15--26.
  Springer (2008)

\bibitem[{Lee and Wong(2001)}]{lee2001elliptic}
Lee, L.p., Wong, K.w.: An elliptic curve random number generator.
\newblock In: Communications and Multimedia Security Issues of the New Century,
  pp. 127--133. Springer (2001)

\bibitem[{Lee et~al.(2020)Lee, Shin, and Hur}]{lee2020return}
Lee, S., Shin, Y., Hur, J.: Return of version downgrade attack in the era of
  tls 1.3.
\newblock In: Proceedings of the 16th International Conference on emerging
  Networking EXperiments and Technologies. pp. 157--168 (2020)

\bibitem[{Liu et~al.(2017)Liu, Zhang, Sun, and Shi}]{liu2017vehicle}
Liu, J., Zhang, S., Sun, W., Shi, Y.: In-vehicle network attacks and
  countermeasures: Challenges and future directions.
\newblock IEEE Network 31(5), 50--58 (2017)

\bibitem[{Luo et~al.(2015)Luo, Wang, Wang, Hu, Cai, and Sun}]{luo2015rpah}
Luo, Y.B., Wang, B.S., Wang, X.F., Hu, X.F., Cai, G.L., Sun, H.: Rpah: Random
  port and address hopping for thwarting internal and external adversaries.
\newblock In: 2015 IEEE Trustcom/BigDataSE/ISPA. vol.~1, pp. 263--270. IEEE
  (2015)

\bibitem[{MacFarland and Shue(2015)}]{macfarland2015sdn}
MacFarland, D.C., Shue, C.A.: The sdn shuffle: creating a moving-target defense
  using host-based software-defined networking.
\newblock In: Proceedings of the Second ACM Workshop on Moving Target Defense.
  pp. 37--41 (2015)

\bibitem[{Mannai et~al.(2018)Mannai, Becheikh, and Rhouma}]{mannai2018new}
Mannai, O., Becheikh, R., Rhouma, R.: A new stream cipher based on nonlinear
  dynamic system.
\newblock In: 2018 26th European Signal Processing Conference (EUSIPCO). pp.
  316--320. IEEE (2018)

\bibitem[{Maurer(1991)}]{maurer1991new}
Maurer, U.M.: New approaches to the design of self-synchronizing stream
  ciphers.
\newblock In: Workshop on the Theory and Application of of Cryptographic
  Techniques. pp. 458--471. Springer (1991)

\bibitem[{Meneghello et~al.(2019)Meneghello, Calore, Zucchetto, Polese, and
  Zanella}]{meneghello2019iot}
Meneghello, F., Calore, M., Zucchetto, D., Polese, M., Zanella, A.: Iot:
  Internet of threats? a survey of practical security vulnerabilities in real
  iot devices.
\newblock IEEE Internet of Things Journal 6(5), 8182--8201 (2019)

\bibitem[{Mill{\'e}rioux and Guillot(2010)}]{millerioux2010self}
Mill{\'e}rioux, G., Guillot, P.: Self-synchronizing stream ciphers and
  dynamical systems: state of the art and open issues.
\newblock International Journal of Bifurcation and Chaos 20(09), 2979--2991
  (2010)

\bibitem[{Nebbione and Calzarossa(2020)}]{nebbione2020security}
Nebbione, G., Calzarossa, M.C.: Security of iot application layer protocols:
  Challenges and findings.
\newblock Future Internet 12(3), 55 (2020)

\bibitem[{NITRD(2013)}]{nitrd2013iwg}
NITRD, C.: Iwg: Cybersecurity game-change research and development
  recommendations (2013)

\bibitem[{Okhravi et~al.(2012)Okhravi, Comella, Robinson, and
  Haines}]{okhravi2012creating}
Okhravi, H., Comella, A., Robinson, E., Haines, J.: Creating a cyber moving
  target for critical infrastructure applications using platform diversity.
\newblock International Journal of Critical Infrastructure Protection 5(1),
  30--39 (2012)

\bibitem[{Park et~al.(2020)Park, Lee, Kang, Lee, and Park}]{park2020ghost}
Park, J.G., Lee, Y., Kang, K.W., Lee, S.H., Park, K.W.: Ghost-mtd: Moving
  target defense via protocol mutation for mission-critical cloud systems.
\newblock Energies 13(8), 1883 (2020)

\bibitem[{Paul and Maitra(2011)}]{paul2011rc4}
Paul, G., Maitra, S.: RC4 stream cipher and its variants.
\newblock CRC press (2011)

\bibitem[{Salmon et~al.(2011)Salmon, Moraes, Dror, and
  Shaw}]{salmon2011parallel}
Salmon, J.K., Moraes, M.A., Dror, R.O., Shaw, D.E.: Parallel random numbers: as
  easy as 1, 2, 3.
\newblock In: Proceedings of 2011 International Conference for High Performance
  Computing, Networking, Storage and Analysis. pp. 1--12 (2011)

\bibitem[{Seibert et~al.(2014)Seibert, Okhravi, and
  S{\"o}derstr{\"o}m}]{seibert2014information}
Seibert, J., Okhravi, H., S{\"o}derstr{\"o}m, E.: Information leaks without
  memory disclosures: Remote side channel attacks on diversified code.
\newblock In: Proceedings of the 2014 ACM SIGSAC Conference on Computer and
  Communications Security. pp. 54--65 (2014)

\bibitem[{Sjoholmsierchio et~al.(2020)Sjoholmsierchio, Hale, Lukaszewski, and
  Xie}]{sjoholmsierchio2020strengthening}
Sjoholmsierchio, M., Hale, B., Lukaszewski, D., Xie, G.G.: Strengthening sdn
  security: Protocol dialecting and downgrade attacks.
\newblock arXiv preprint arXiv:2010.11870  (2020)

\bibitem[{Vaccari et~al.(2020)Vaccari, Aiello, and
  Cambiaso}]{vaccari2020slowite}
Vaccari, I., Aiello, M., Cambiaso, E.: Slowite, a novel denial of service
  attack affecting mqtt.
\newblock Sensors 20(10), 2932 (2020)

\bibitem[{Vuong et~al.(2015)Vuong, Loukas, Gan, and
  Bezemskij}]{vuong2015decision}
Vuong, T.P., Loukas, G., Gan, D., Bezemskij, A.: Decision tree-based detection
  of denial of service and command injection attacks on robotic vehicles.
\newblock In: 2015 IEEE International Workshop on Information Forensics and
  Security (WIFS). pp. 1--6. IEEE (2015)

\bibitem[{Wang et~al.(2017)Wang, Chen, and Zhu}]{wang2017random}
Wang, K., Chen, X., Zhu, Y.: Random domain name and address mutation (rdam) for
  thwarting reconnaissance attacks.
\newblock PloS one 12(5), e0177111 (2017)

\bibitem[{Wang and Cheng(2019)}]{wang2019pseudo}
Wang, L., Cheng, H.: Pseudo-random number generator based on logistic chaotic
  system.
\newblock Entropy 21(10), 960 (2019)

\end{thebibliography}

\end{document}